\documentclass[a4paper,fleqn,usenatbib]{mnras}
\usepackage{graphicx}
\usepackage{amsmath}
\usepackage{amssymb}
\usepackage{upgreek}
\usepackage{longtable}

\title[Pulsars with interpulse emission]
{On the beam properties of radio pulsars with interpulse emission}

\author[Johnston \& Kramer]
{Simon Johnston$^{1}$\thanks{email: Simon.Johnston@csiro.au} and
Michael Kramer$^{2,3}$\\
\\
$^{1}$CSIRO Astronomy and Space Science, Australia Telescope National Facility, PO Box 76, Epping, NSW 1710, Australia\\
$^{2}$Max-Planck-Institut f\"ur Radioastronomie (MPIfR), Auf dem H\"ugel 69, D-53121 Bonn, Germany\\
$^{3}$University of Manchester, Jodrell Bank Centre for Astrophysics, Alan Turing Building, Manchester M13 9PL\\
}
\date{Accepted \today. Received \today; in original form \today}

\pubyear{2018}

\begin{document}
\label{firstpage}
\pagerange{\pageref{firstpage}--\pageref{lastpage}} 
\maketitle

\begin{abstract}
In the canonical picture of pulsars, radio emission arises from a narrow
cone centered on the star's magnetic axis but many basic details remain unclear. 
We use high-quality polarization data taken with the Parkes radio telescope to constrain the geometry and emission heights of pulsars showing interpulse emission, and include the possibility that emission heights in the main and interpulse may be different. We show that emission heights are low in the centre of the beam, typically less than 3\% of the light cylinder radius. The emission beams are under-filled in longitude, with an average profile width only 60\% of the maximal beam width and there is a strong preference for the visible emission to be located on the trailing part of the beam. We show substantial evidence that the emission heights are larger at the beam edges than in the beam centre. There is some indication that a fan-like emission beam explains the data better than conal structures. Finally, there is a strong correlation between handedness of circular polarization in the main and interpulse profiles which implies that the hand of circular polarization is determined by the hemisphere of the visible emission.
\end{abstract}


\section{Introduction}
Understanding the properties of the emission beams of radio pulsars has a number of consequences for a variety of questions. The fraction of the actual illuminated sky impacts on estimates on how many radio emitting neutron stars are in the Galaxy. This, in turn, should be compared to the core collapse supernova rate, believed to the birth events of pulsars \citep{kk08}. The structure of the emission beam is likely to be affected -- or even determined -- by the still unknown underlying emission mechanism, the heights at which the emission originates and/or leaves the magnetosphere. The size of the beam is affected by the spin period. The inclination of the emission beam with respect to the rotation axis, and whether it changes with time, determines the evolution and spin-down behaviour of the neutron star \citep{jk17}. Finally, the radio beam may be related (e.g. in phase and/or origin) to the emission seen across the electromagnetic spectrum. 

In this context, the polarization properties of radio pulsars can be a most powerful tool for understanding both the geometry of the star and the emission properties (e.g. \citealt{lm88,ran90,ew01,dgm+15,dkl+19}). In this paper we will use the polarization properties to examine five major questions:
\begin{itemize}
\item[-] How filled are the radio beams in the longitudinal direction?
\item[-] Is the emission height lower in the centre of the beam and higher towards the edges \citep{gg03,kj07,dkl+19}?
\item[-] Can we distinguish between models of the beam which are conal \citep{ran93,mr02} versus fan-like \citep{wpz+14,dr15}?
\item[-] Does the observed profile and its polarization properties depend on the geometry \citep{bes18}?
\item[-] Is there a relationship between properties of the linear and circular polarisation that reflects geometry?
\end{itemize}
To answer these questions, we concentrate on pulsars which have ``interpulse" emission and are therefore close to orthogonal rotators with $\alpha$, the angle between the magnetic and rotation axis, near 90\degr, so that emission from both poles be observed. Apart from observing two (related) cuts through the two poles' emssion beams, the existence of polarised emission nearly 180\degr\ apart in rotational phase, makes the determination of geometry unusually reliable and precise. Furthermore, previous work on interpulse pulsars \citep{wj08a,kj08,kjwk10,mgr11,dkc+13,abp17,dkl+19} have given partial answers to the questions posed above, but here we develop a new technique for determining the absolute emission heights for both main and interpulse emission which enable us to construct a beam map for the location of the radio emission.

In the rotating vector model (RVM) of \citet{rc69}, the radiation is beamed along the field lines and the plane of polarization is determined by the angle of the magnetic field as it sweeps past the line of sight. The position angle (PA) as a function of pulse longitude, $\phi$, can be expressed as
\begin{equation}
\label{paswing}
{\rm PA} = {\rm PA}_{0} +
{\rm arctan} \left( \frac{{\rm sin}\alpha
\, {\rm sin}(\phi - \phi_0)}{{\rm sin}\zeta
\, {\rm cos}\alpha - {\rm cos}\zeta
\, {\rm sin}\alpha \, {\rm cos}(\phi - \phi_0)} \right)
\end{equation}
Here, $\alpha$ is the angle between the rotation axis and the magnetic axis and $\zeta=\alpha+\beta$ with $\beta$ being the angle of closest approach of the line of sight to the magnetic axis. $\phi_0$ is the pulse longitude at which the PA is PA$_{0}$, which also corresponds to the PA of the rotation axis projected onto the plane of the sky. The best evidence for the validity of the RVM and the notion that the PA swing reflects the geometry of the pulsar was recently given by \citet{dkl+19}.

If the radio emission occurs at a finite height above the pulsar surface, the PA traverse is expected to be delayed with respect to the total intensity profile by relativistic effects such as aberration and retardation (hereafter referred to as A/R) as initially discussed by \citet{bcw91}. Defining the radius of the light cylinder, $R_{lc}$, as
\begin{equation}
R_{lc} = \frac{P\,\,c}{2\,\,\pi}
\end{equation}
where $P$ is the pulsar period and $c$ the speed of light then the magnitude of the shift in longitude, $\delta\phi$(PA) (in radians), is related to the emission height relative to the centre of the star, $h_{\rm em}$, via
\begin{equation}
\label{bcw}
\delta\phi({\rm PA}) = \frac{4\,\,h_{\rm em}}{R_{lc}}
\end{equation}
\citep{bcw91}.
This allows the emission height to be derived if one can determine the location of the magnetic pole within the pulse profile and can determine the location of PA$_{0}$. Although the fact that $\delta\phi({\rm PA}) \propto 4h_{\rm em}$ is agreed upon, according to \citet{bcw91} the total intensity profile is shifted earlier by an amount proportional to $h_{\rm em}$ and the PA swing is shifted to later times by $3h_{\rm em}$. In contrast, in \citet{dyks08} the shift is 2$h_{\rm em}$ for both the intensity profile and the PA swing. The results presented recently by \citet{dkl+19} suggest that most or all the shift occurs in the total intensity rather than the PA swing. Earlier, \citet{cr12} showed that computation of $h_{em}$ also needs to be modified especially when $h_{em}$ is larger than about $0.1 R_{lc}$.

Once the emission height is known, simple geometry (and the assumption of dipolar field lines) allows the calculation of the half-opening angle of the emission cone at the last open field line, $\rho_o$, via
\begin{equation}
\label{height}
\rho_o = \sqrt{\frac{9\,\, h_{\rm em}}{4\,\,R_{lc}}}
\end{equation}
\citep{ran90}.
The combination of $\rho_o$, $\alpha$ and $\beta$, then informs us of the maximum observable pulse width. When comparing this with data, we typically choose to measure the width at a 10\% intensity level, $W_{10}$. Assuming a filled circular emission zone, 
\begin{equation}
\label{rho}
{\rm cos}\rho_e = {\rm cos}\alpha\,\, {\rm cos}\zeta\,\, +\,\, {\rm sin}\alpha\,\, {\rm sin}\zeta\,\, {\rm cos}(W_{10}/2)
\end{equation}
\citep{ggr84}. Measurement of the actual pulse width then gives an indication of the filling factor of the beam \citep{jk19}.

\begin{table}
\caption{Pulsar period, $P$, and widths, $W_{10}$, for the main and interpulse.
$\rho$ and $h$ for the main and interpulse are computed from
Equations~\ref{rho} and \ref{height}.}
\label{psrinfo}
\resizebox{0.50\textwidth}{!}{
\begin{tabular}{lrrrrrrrr}
PSR (J2000) & $P$ & $r_{lc}$ & $W_{10M}$ & $\rho_M$ & $h_M$ &$W_{10I}$ & $\rho_I$ & $h_I$\\
    & (ms) & ($10^3$km) & (deg) & (deg) & (km) & (deg) & (deg) & (km) \\
\hline & \vspace{-3mm} \\
0627$+$0705   & 476 & 22.7 & 6.7 & 11.2 & 380 & 14.4 &  7.2 & 160 \\
0905$-$5127  & 346 & 16.5 &11.6 & 20.9 & 970 &  8.0 &  4.0 &  40 \\
0908$-$4913  & 107 &  5.1 &12.3 &  9.6 &  60 & 16.9 & 10.0 &  70 \\
1549$-$4848  & 288 & 13.7 &15.4 &  8.3 & 130 & 15.0 &  7.6 & 110 \\
1611$-$5209  & 182 &  8.7 & 5.6 &  5.9 &  40 & 10.0 &  8.5 &  90 \\
1637$-$4553  & 119 &  5.7 &18.3 &  9.1 &  60 & 18.0 & 10.1 &  80 \\
1705$-$1906  & 299 & 14.3 &16.5 & 11.6 & 260 &  5.9 &  3.5 &  20 \\
1722$-$3712  & 236 & 11.3 &13.4 &  8.7 & 110 & 17.9 & 14.0 & 300 \\
1739$-$2903  & 323 & 15.4 &10.9 & 10.6 & 240 & 14.8 &  9.8 & 200 \\
1828$-$1101  & 72 &  3.4 & 7.7 &  8.6 &  35 & 13.0 &  9.7 &  45 \\
1935$+$2025   & 80 &  3.8 &22.1 & 19.8 & 200 & 12.0 &  7.9 &  30 \\
\end{tabular}}
\end{table}

The finite height of the emission means some alterations to Equation~\ref{paswing} need to be made. \citet{ha01} showed that the PA swing is shifted in a vertical direction by an amount proportional to ${\rm cos}(\alpha)$. \citet{dyks08} showed how to deal with the case where $h_{em}$ varied across the polar cap (i.e. as a function of $\phi$). The modified  Equation~\ref{paswing} looks as follows:
\begin{equation}
\label{paswing_mod}
\begin{split}
{\rm PA} & = {\rm PA}_{0} + \\
& {\rm arctan} \left( \frac{{\rm sin}\alpha
\, {\rm sin}(\phi - \phi_0 - 2h_{em}/R_{lc})}{{\rm sin}\zeta
\, {\rm cos}\alpha - {\rm cos}\zeta
\, {\rm sin}\alpha \, {\rm cos}(\phi - \phi_0 - 2h_{em}/R_{lc})} \right) \\
& + \frac{10}{3}\frac{h_{em}}{R_{lc}}{\rm cos}\alpha
\end{split}
\end{equation}

Finally, in the case of interpulse pulsars, there is no {\it a priori} reason to expect that the height of the main pulse emission should be the same as that of the interpulse emission, especially if the values of $\beta$ are significantly different. We therefore modify Equation~\ref{paswing} by adding in a term $\Delta$ which represents the difference in emission height between the two poles as follows:
\begin{equation}
\label{paswing2}
{\rm PA} = {\rm PA}_{0} +
{\rm arctan} \left( \frac{{\rm sin}\alpha
\, {\rm sin}(\phi - \phi_0 - \Delta)}{{\rm sin}\zeta
\, {\rm cos}\alpha - {\rm cos}\zeta
\, {\rm sin}\alpha \, {\rm cos}(\phi - \phi_0 - \Delta)} \right)
\end{equation}
where we ignore the final term of Equation~\ref{paswing_mod} because ${\rm cos}\alpha$ is small for orthogonal rotators.
As written, a positive value of $\Delta$ implies a lower emission height of the main pulse compared to the interpulse.

A further constraint on emission heights comes via the work of \citet{gan04} and \citet{ym14}.  In those papers the authors point out that, assuming the emission originates along the tangents to the dipolar field lines, purely  geometrical effects mean that the radio emission must have a minimum height below which it cannot be seen by an Earth-based observer. This minimum height depends on $\alpha$ and $\zeta$ and increases as the line of sight moves further from the magnetic axis.

The plan of the paper is as follows. After describing the observations and data analysis, we present the results for each pulsar, including the derived geometry. We then discuss the sample of studied pulsars as a whole, before commenting on the individual sources. This leads to implications related to the five questions posed above, which are summarised and discussed. 

\section{Observations and Results}
Observations at 1.4~GHz were obtained from the recent compilation of pulsar polarization profiles presented in \citet{jk18}. Details of the flux density and polarization calibration can be found in that paper. Table~\ref{psrinfo} gives the pulsar period and the measured pulse widths at the 10\% level for the main and interpulse profiles taken from \citet{jk19}. From these widths, $\rho$ and $h$ can then be determined from Equations~\ref{rho} and \ref{height}, assuming that the observed pulse width is centred on the magnetic axis. Consequently, these can be viewed as the minimum possible values in order to ensure that the emission fills the beam and is the value generally quoted in the literature. However, the beam can be under-filled, and the computation does not taken into account the effects of retardation and aberration. We will account for those possibilities later.
\begin{figure}
\includegraphics[width=0.45\textwidth]{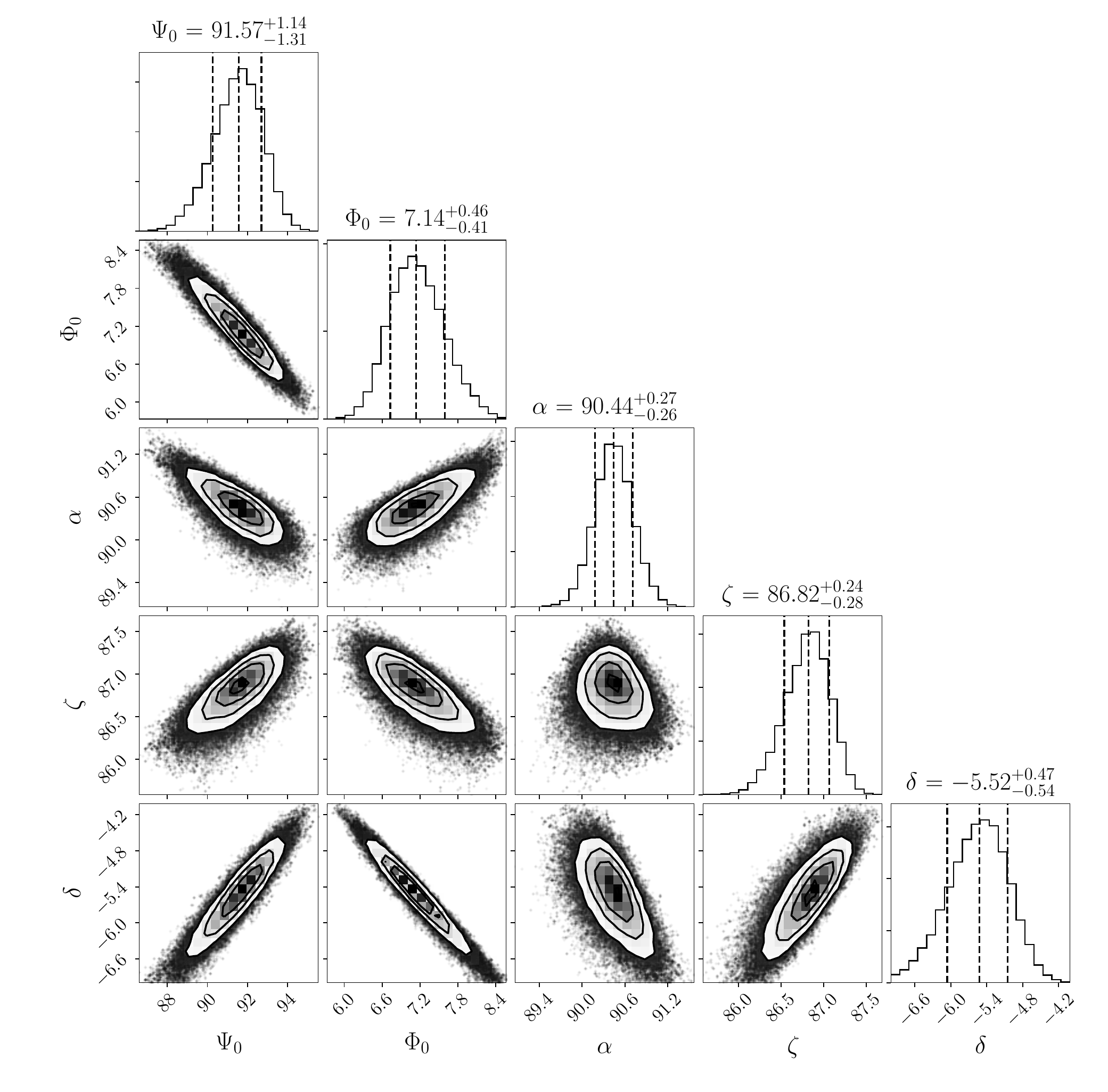}
\caption{Posterior distributions of the five parameters included in our RVM fits as resulting from MCMC for PSR J0905$-$5157. The top of each column shows the one-dimensional posterior for each parameter, while the other plots show the corresponding correlations between them. The dashed vertical lines indicate the median and the 16 and 84 percentiles of the distribution, respectively. The units of all angles is degrees.}
\label{corner}
\end{figure}

\subsection{Determining Rotating Vector Models}
There are five unknowns in Equation~\ref{paswing2}; $\alpha$, $\zeta$, PA$_0$, $\phi_0$ and $\Delta$. In order to determine their values for each pulsar, we performed Markov Chain Monte Carlo (MCMC) fits, using the implementation of Goodman \& Weare’s Affine Invariant MCMC Ensemble sampler as implemented in the python package EMCEE \citep{fhlg13}. We set uniform priors for angles and phases between 0 and 360\degr, and for $\Delta$ between $-15$\degr\ and $+15$\degr, with the rationale that uniform priors are non-informative, and offer no biases to our posterior estimates.

\begin{table*}
\caption{Results of fitting the RVM model. $\alpha_{M}$, $\beta_{M}$
and $\phi_{M}$ refer to the main pulse, $\alpha_{I}$, $\beta_{I}$
and $\phi_{I}$ to the interpulse.
$\phi$ is the location of the inflexion point of the PA swing with respect
to the peak of the main pulse emission. All angles in degrees.}
\label{bigtable}
\begin{tabular}{lrrrrrrrrr}
Jname & $\alpha_{M}$ & $\beta_{M}$ & $\phi_{M}$ & $\zeta$ & $\alpha_{I}$ & $\beta_{I}$ &
$\phi_{I}$ & $\Delta$ & $h_M - h_I$\\
\hline & \vspace{-3mm} \\
0627$+$0705  & 94.0(2)  & --8.6 &   4.7(2)  &    85.4(2)  &    86.0(2) & --0.6  &   184.8    &    0.1(2)  &   10(20)    \\
0905$-$5157 & 90.4(3)  & --3.6 &   7.1(5)  &    86.8(3)  &    89.5(3) & --2.8  &   181.6(5) &   --5.5(5) &  400(40)    \\
0908$-$4913 & 83.59(4) & 7.44  &   3.7(1)  &    91.03(7) &    96.41(4) & --5.38 &   183.41  &    0.29(7) &    --10(2) \\
1549$-$4848 & 87.3(3)  & 2.3   &    0.9(3) &   89.5(1.0) &    92.7(3) & --3.2  &   183.1(3) &     2.2(5) &  --130(30) \\
1722$-$3712 & 87.4(5)  & --5.6 &    4.7(1) &    81.7(6)  &    92.6(5) & --10.9 &   188.1(1) &     3.5(9) &  --170(40) \\
1739$-$2903 & 94.6(2)  & --2.2 &  --2.8(1) &    92.4(1)  &    85.4(2) &  7.1   &   181.1(1) &    3.9(3) & --260(20) \\
1828$-$1101 & 82.7(2)  & 7.3   &  --1.8(8) &    90.0(2) &    97.3(2) &  --7.3  &   179.6(8) &    1.4(1.6) &  20(25) \\
1935$+$2025  & 93(2) & --13 & --13(2) &  80(1) &  87(2) & --7 &   166(1) &  --1(1) &   20(20) \\
\end{tabular}
\end{table*}

\begin{table*}
\caption{Parameters for $\rho$ and $h$ for the main and interpulses
for 8 pulsars in order to ensure that the emission is located within
the open field lines. The filling fraction is also listed.}
\label{beamtab}
\begin{tabular}{lrrrrrrrrrr}
Jname & $\rho_M$ & $h_{M}$ (km) & $h_M/r_{lc}$ & fill
      & $\rho_I$ & $h_{I}$ (km) & $h_I/r_{lc}$ & fill\\
\hline & \vspace{-3mm} \\
0627$+$0705   &  9.9 -- 24.2 &  270 -- 1800 & 0.012 -- 0.079 & 0.68 -- 0.15 &  9.5 -- 24.2 & 280 -- 1810 & 0.012 -- 0.080 & 0.76 -- 0.30 \\
0905$-$5157  & 16.0 -- 28.6 &  580 -- 1700 & 0.035 -- 0.103 & 0.37 -- 0.20 &  9.0 -- 24.2 & 180 -- 1300 & 0.011 -- 0.079 & 0.47 -- 0.17  \\
0908$-$4913  & 13.7 -- 30.6 &  130 --  650 & 0.027 -- 0.127 & 0.53 -- 0.21 & 14.2 -- 30.8 & 140 --  660  & 0.029 -- 0.129 & 0.64 -- 0.28 \\
1549$-$4848  & 11.6 -- 21.1 &  250 -- 830 & 0.018 -- 0.060 & 0.68 -- 0.37 & 14.3 -- 22.7 & 380 -- 960 & 0.028 -- 0.070 & 0.54 -- 0.33 \\
1722$-$3712  & 10.5 -- 29.7 &  170 -- 1350 & 0.015 -- 0.120 & 0.76 -- 0.23 & 15.0 -- 31.6 & 340 -- 1520 & 0.030 -- 0.135 & 0.69 -- 0.24 \\
1739$-$2903  & 10.5 -- 22.2 &  230 -- 1030 & 0.015 -- 0.067 & 0.58 -- 0.25 & 15.3 -- 24.9 & 490 -- 1290 & 0.032 -- 0.084 & 0.51 -- 0.30 \\
1828$-$1101  & 13.9 -- 18.0 &   90 --  150 & 0.026 -- 0.044 & 0.33 -- 0.23 & 16.0 -- 19.6 & 120 -- 180 & 0.035 -- 0.052 & 0.46 -- 0.36 \\
1935$+$2025   & 19.7 &  200 & 0.053 & 0.88 & 20.6 & 220 & 0.058 & 0.30 \\
\end{tabular}
\end{table*}

\begin{figure*}
\begin{tabular}{cc}
\includegraphics[width=0.45\textwidth]{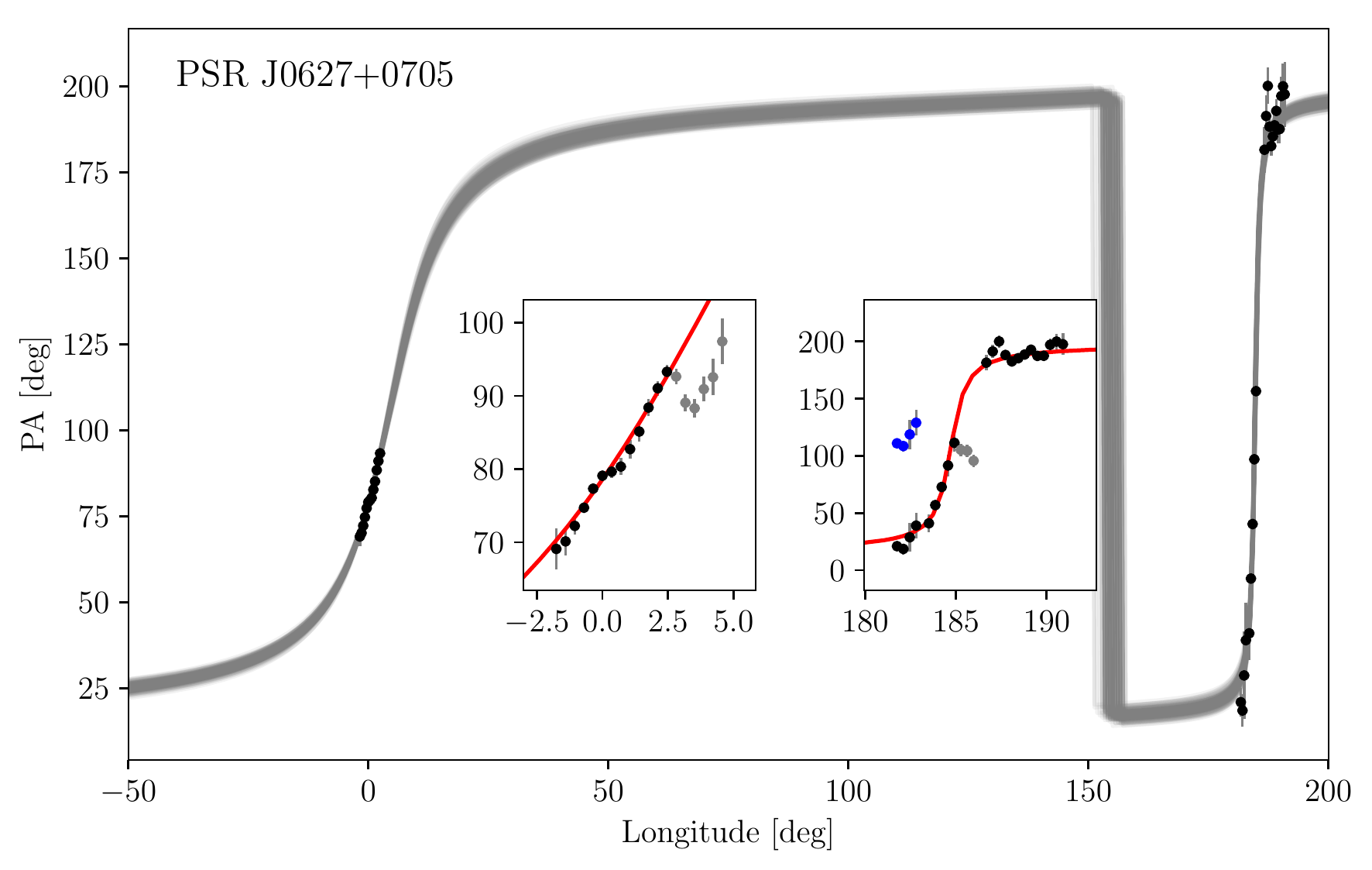} & \includegraphics[width=0.45\textwidth]{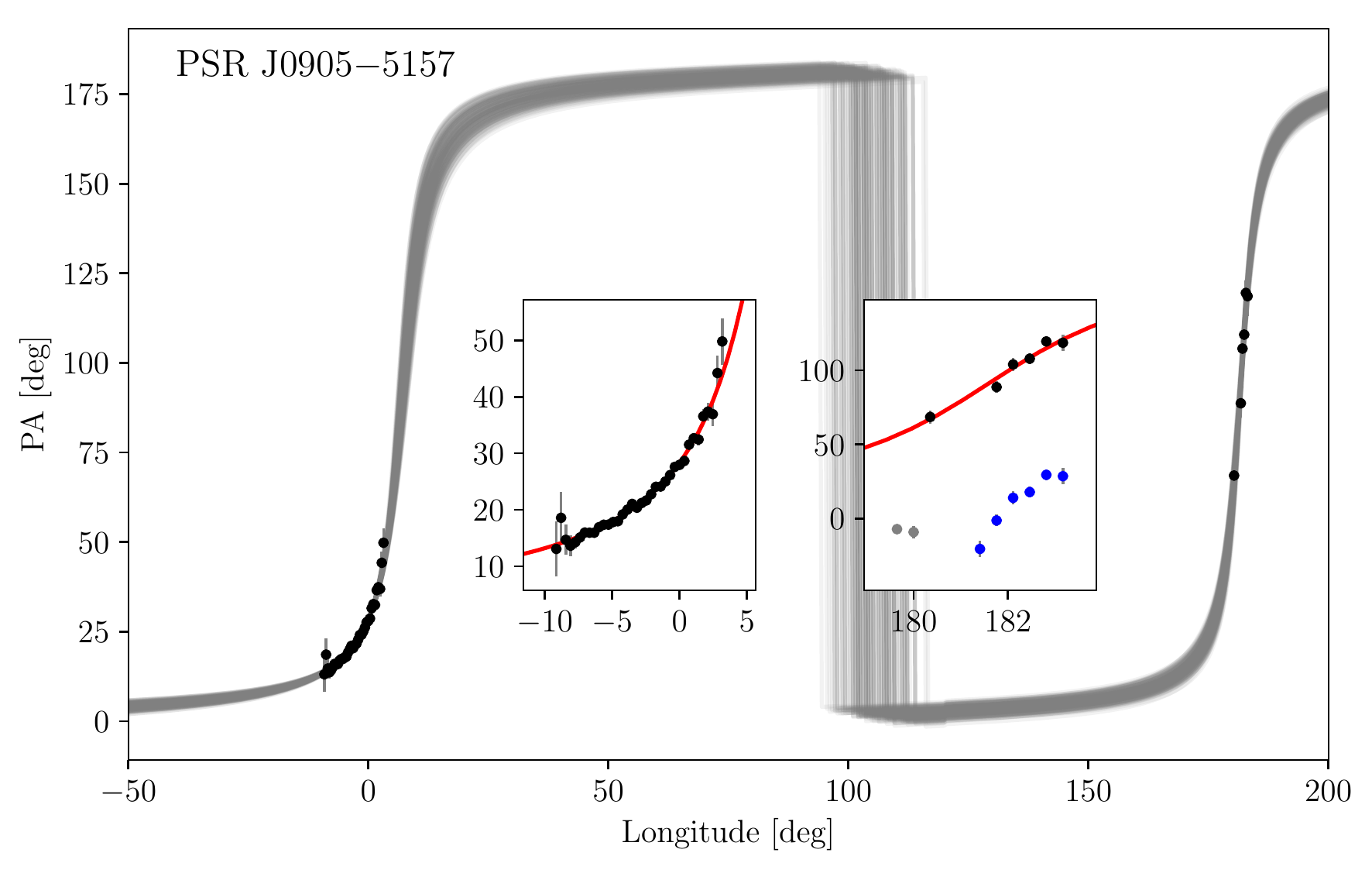} \\
\includegraphics[width=0.45\textwidth]{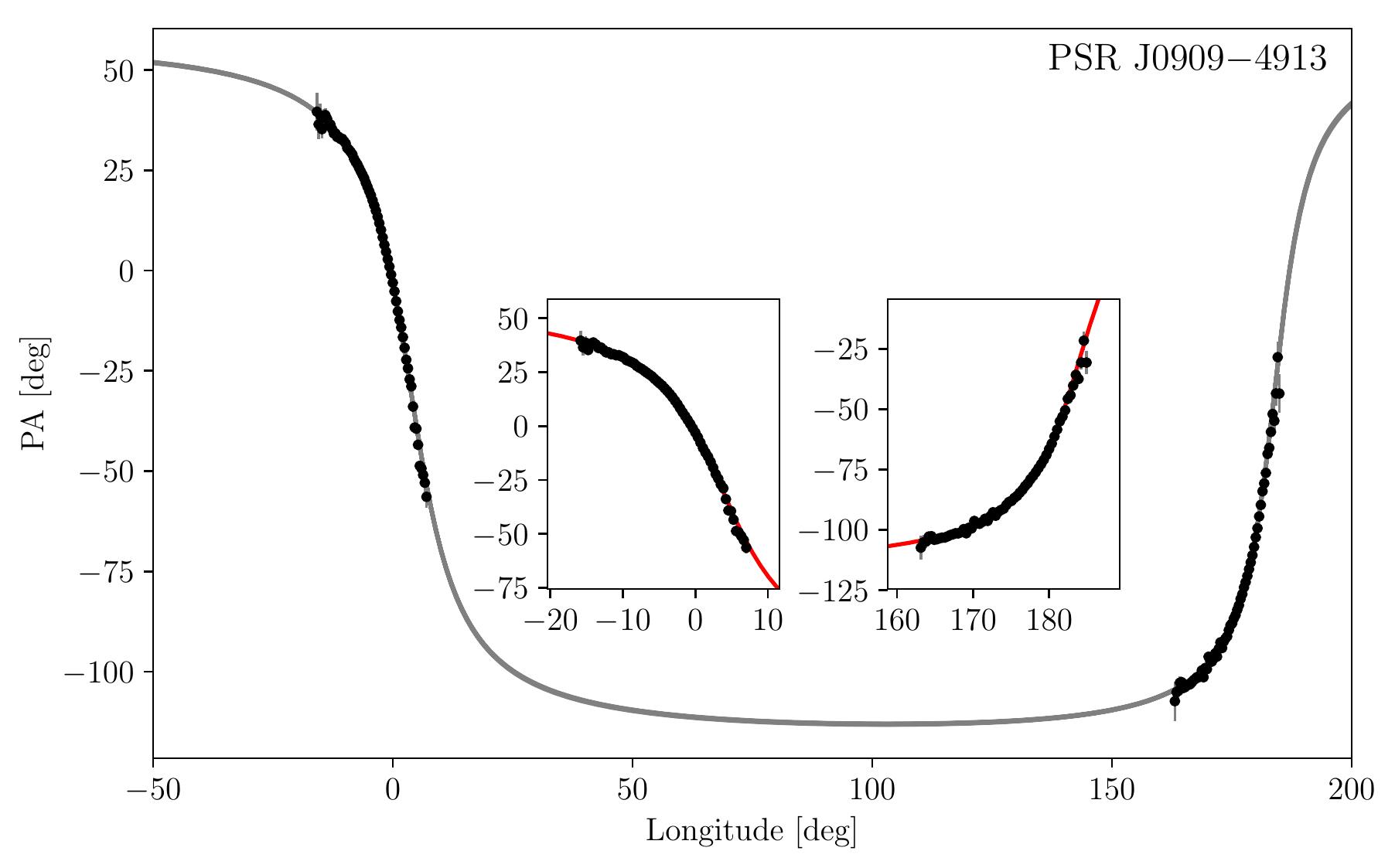} & \includegraphics[width=0.45\textwidth]{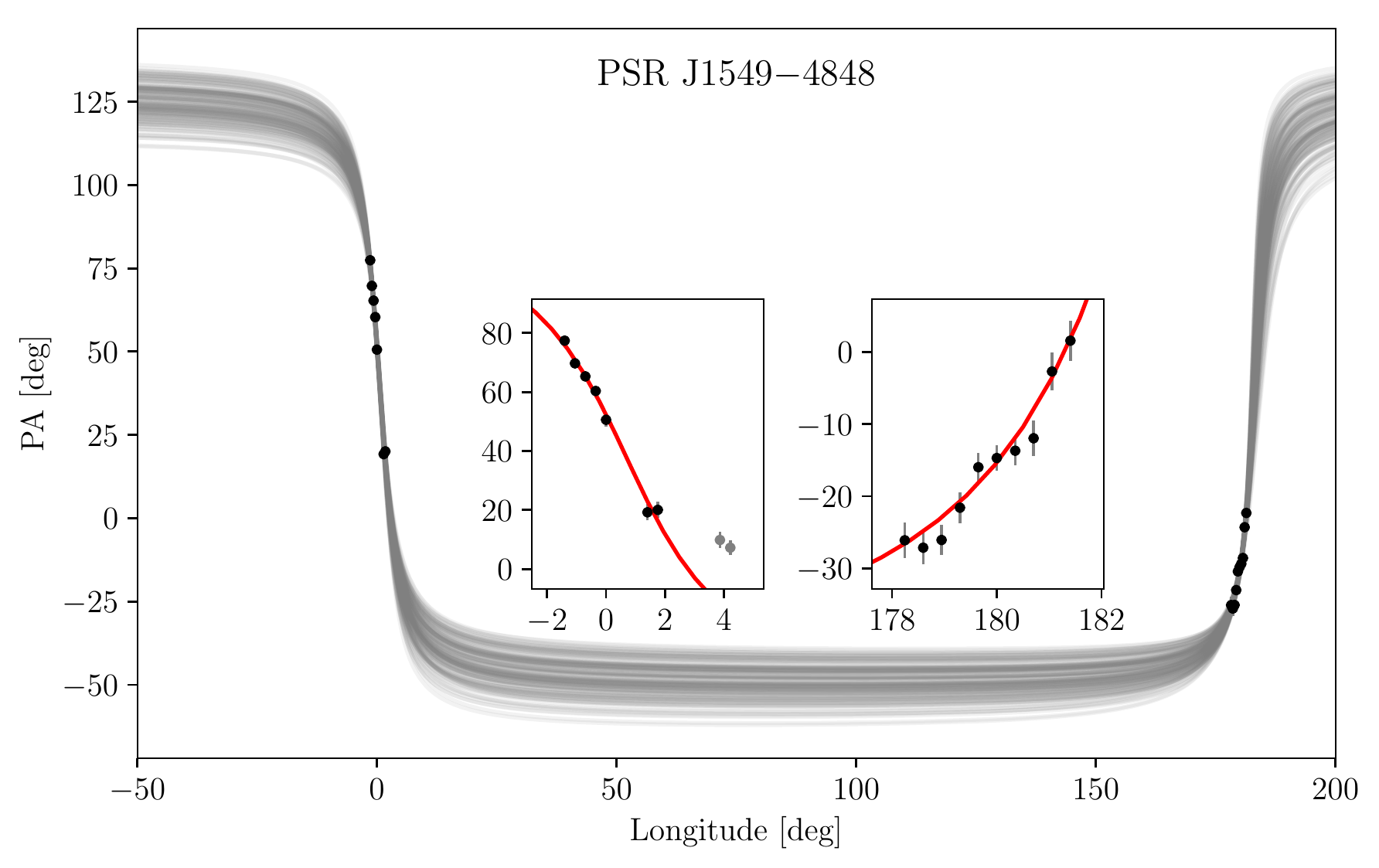} \\
\includegraphics[width=0.45\textwidth]{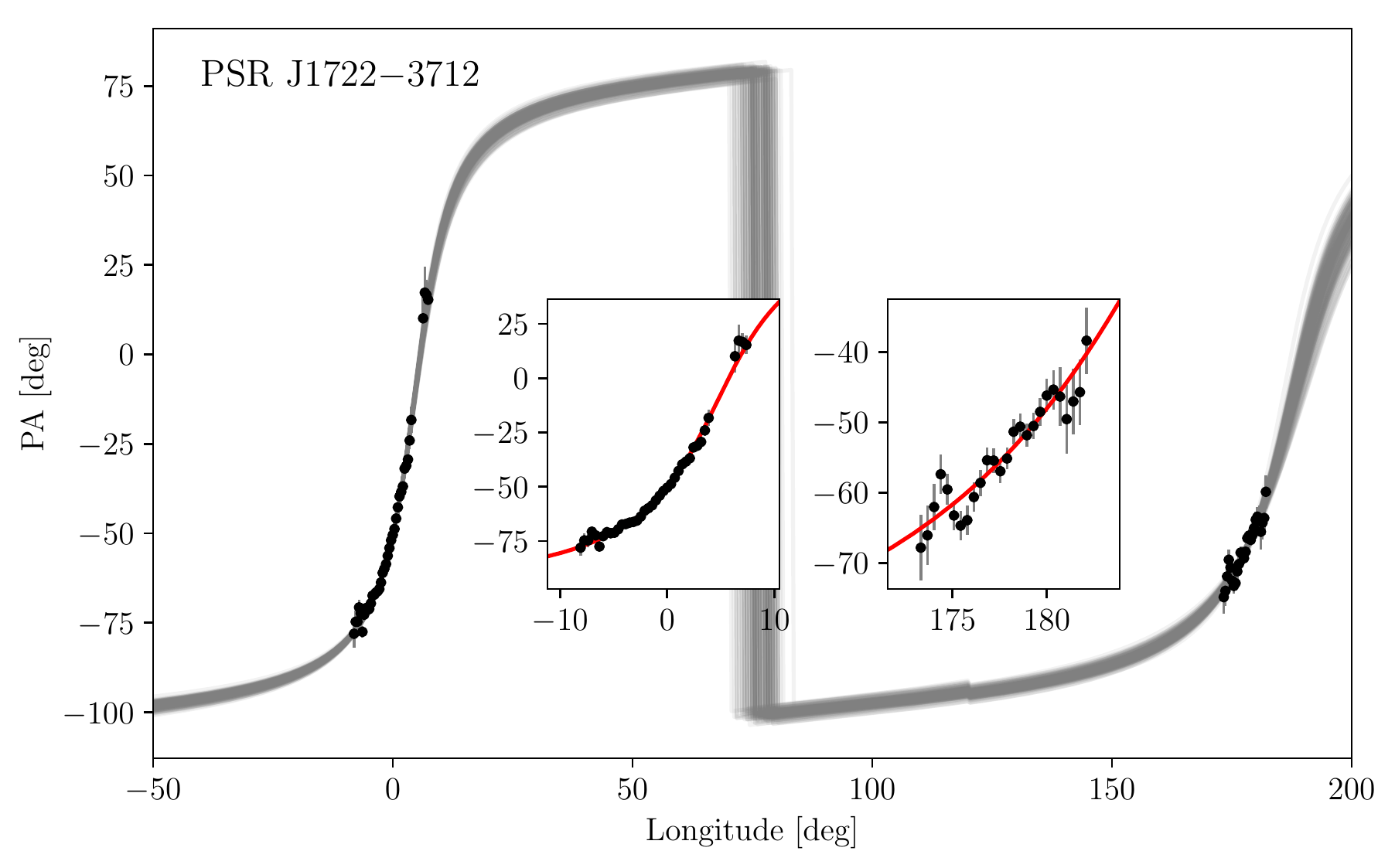} & \includegraphics[width=0.45\textwidth]{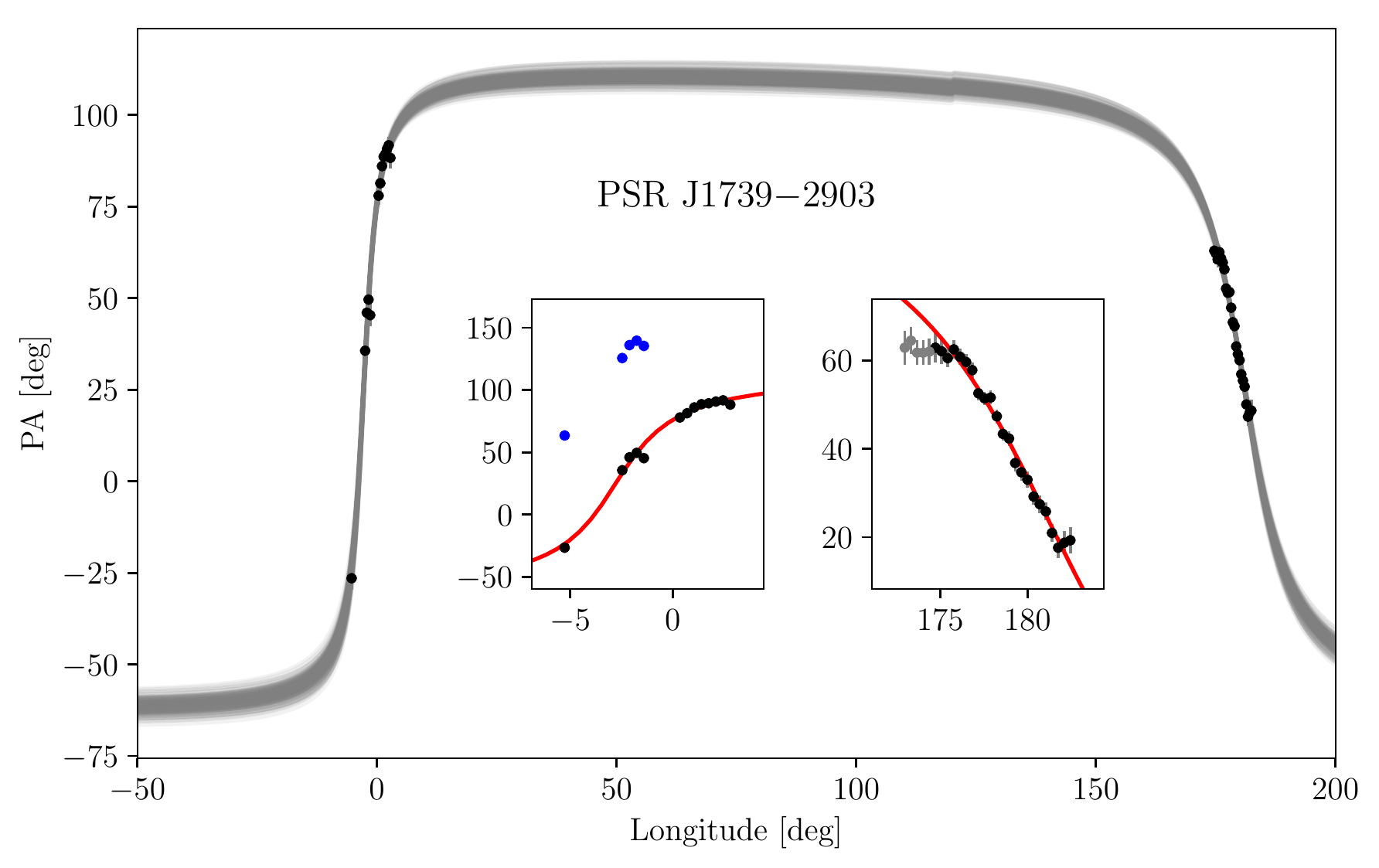} \\
\includegraphics[width=0.45\textwidth]{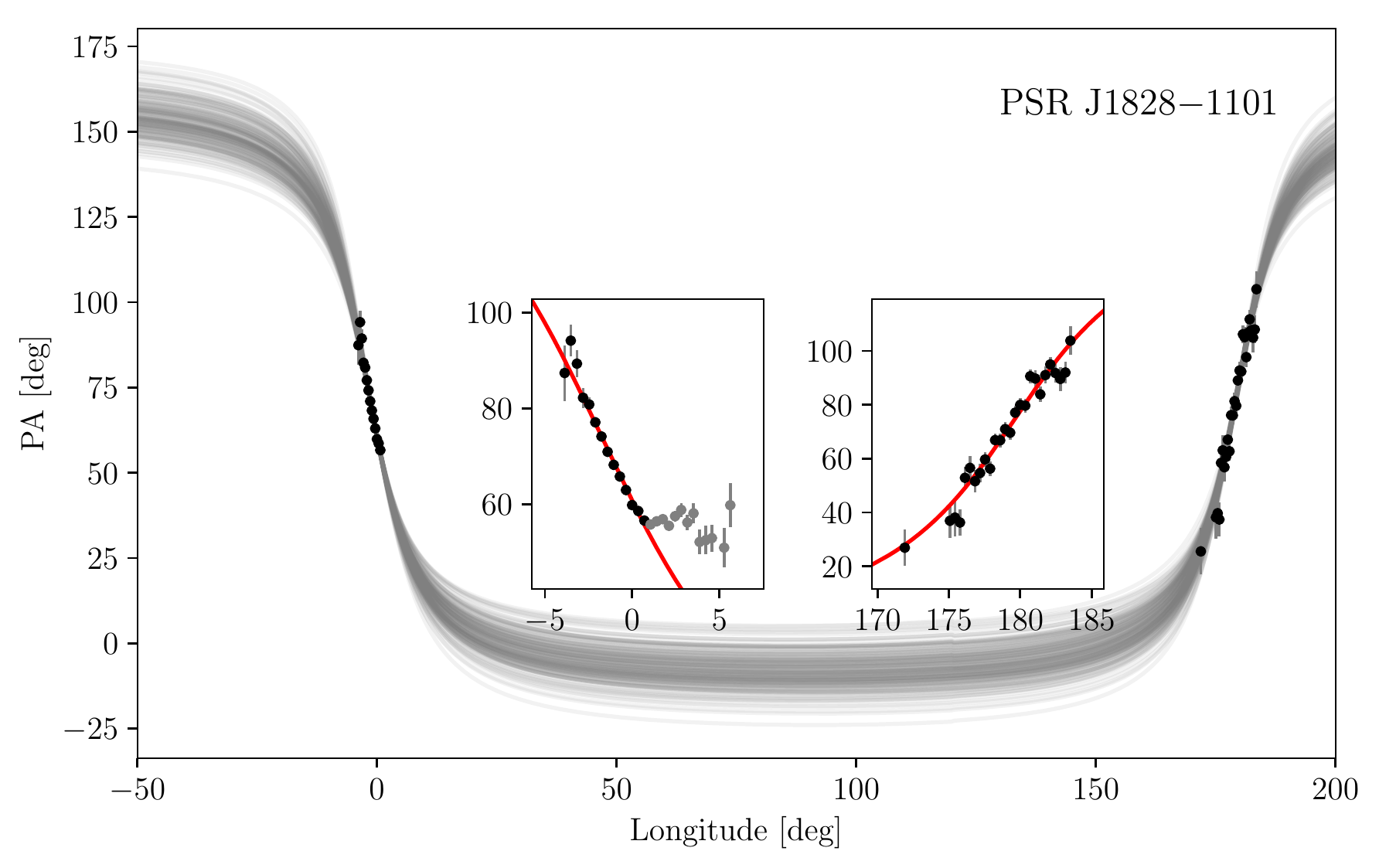} & \includegraphics[width=0.45\textwidth]{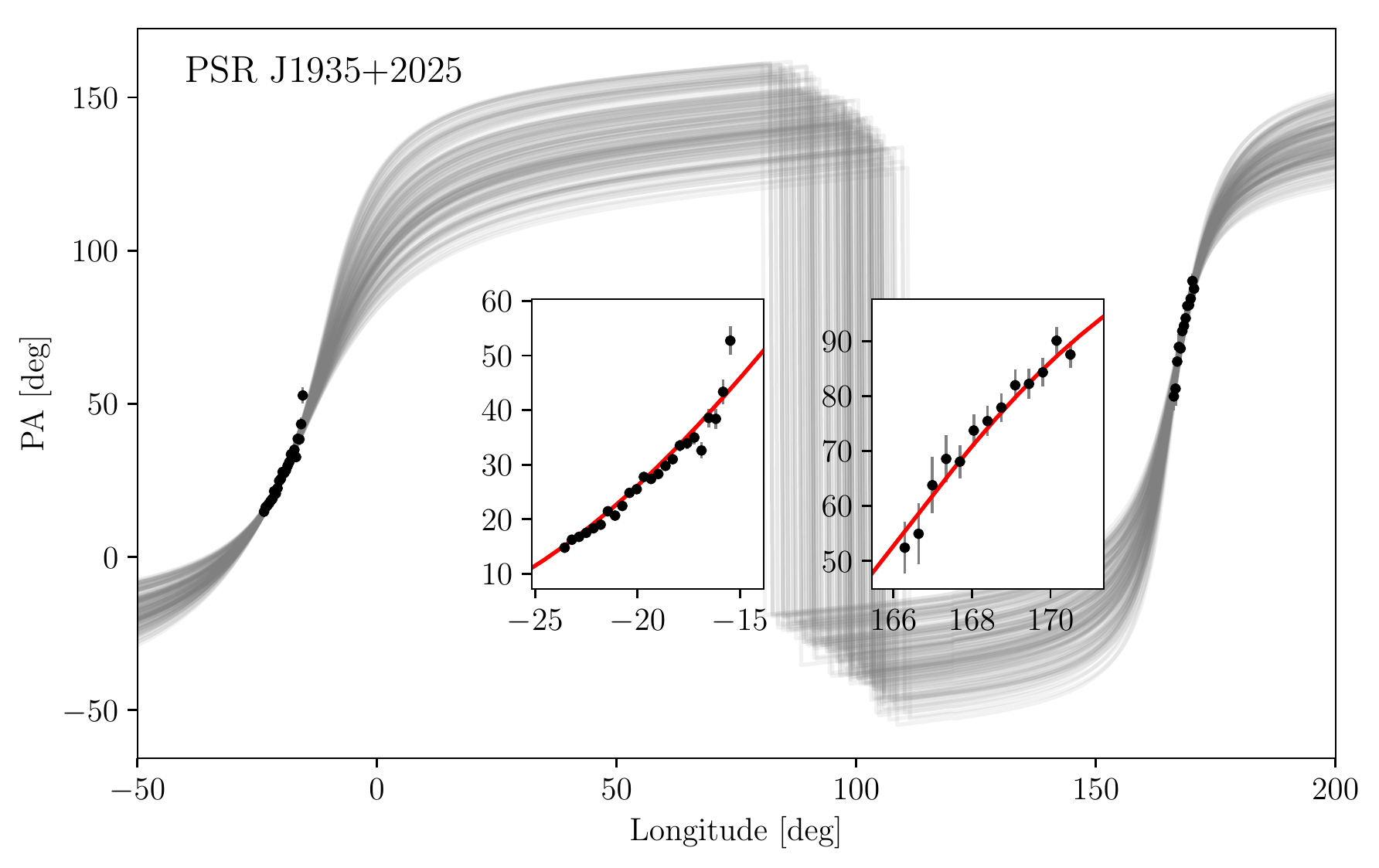} \\
\end{tabular}
\caption{PA swings for 8 pulsars. Each panel shows the observed PA points (black dots) and the PA swing across the entire pulse longitude using the error bars determined by the fitting process. The inset is zoomed in on the main and interpulse. Grey dots are points not included in the fits.  Blue dots denote points with a 90\degr\ phase jump applied.}
\label{pafig}
\end{figure*}

Results of the RVM fitting are given in Table~\ref{bigtable}.
The values reported are the medians of the posterior probability distributions and a 1-$\sigma$ quantile, estimated as half the difference between the 16 and 84 percentiles.
An example `corner' plot for PSR~J0905$-$5157 is given in Figure~\ref{corner}. PA solutions are given in Figure~\ref{pafig}, where we have drawn sets of 150 samples from the posterior distributions, for each plotting the corresponding RVM curve in grey to indicate the uncertainties reflected by the
derived posteriors. The RVM corresponding to the median values is plotted in red in the two insets, which are centred on the main pulse and interpulse, respectively. PA values included in the fit are shown in black. In some cases, we have chosen to ignore certain PA values, even though they nominally fulfill our threshold criteria for significant linear polarised emission. Also, for a number of PA values we have identified them as being emitted in an orthogonal emission mode \citep{brc76}. In those cases, we have applied a 90\degr\ shift. Those data points are indicated blue at their original value. We justify those made choices below, when we discuss the individual pulsars. The spread in grey RVM lines allows us to easily identify those RVM fits and parameters that are better constrained than others.  

As discussed in the previous section the RVM fits yield the relative height difference between the main and interpulse emission, given in the last column of the table. We were unable to obtain constraining fits for PSRs J1611--5209, J1637--4553 or J1705--1906. For PSRs~J1611--5209 and J1637--4553 the interpulse is very weak and the polarization levels are marginal. In addition for the latter pulsar the main pulse has a peculiar position angle swing. For PSR~J1705--1906 the interpulse has a U shaped position angle swing, making it difficult to determine a unique solution (see \citealt{wws07} for a discussion).

\subsection{Emission beams and altitudes}
One piece of the puzzle remains; although we have the {\it relative} difference in height between the main and interpulse emission, we still need to determine the {\it absolute} height. There are two possible ways to proceed. We could locate the location of the magnetic axis on the pulse profile and then use the value of $\phi_0$ and equation~\ref{bcw} to determine the height. This is the method employed in the seminal paper by \citet{bcw91} and others since \citep{jhv+05,ran07,wj08b,nsk+13}. However, there is subjectivity about determining the location of the magnetic axis.  In addition, as described above, it is unclear whether the aberration shift is equal (but opposite in sign) for the total intensity and the polarization swing or not.

Instead, we first assume only that the pulse profile cannot extend beyond the edges of the emission cone \citep{rwj15b} and that the total shift between PA and profile is given by Equation~\ref{bcw}. In this case there is a minimum absolute emission height for which both main and interpulses fit within the beam given the relative height offsets. As the height increases the profile shifts rightwards to compensate the effects of aberration and retardation, so in turn this sets a maximum height when either the main or interpulse profile touches the trailing edge of the cone. Results are given in Table~\ref{beamtab} with the range of permissible heights listed. Note however, that the heights of the main pulse and interpulse are linked via $\Delta$; a given $h_M$ implies fixed $h_I$.

\begin{figure*}
\begin{tabular}{cc}
\includegraphics[width=0.40\textwidth]{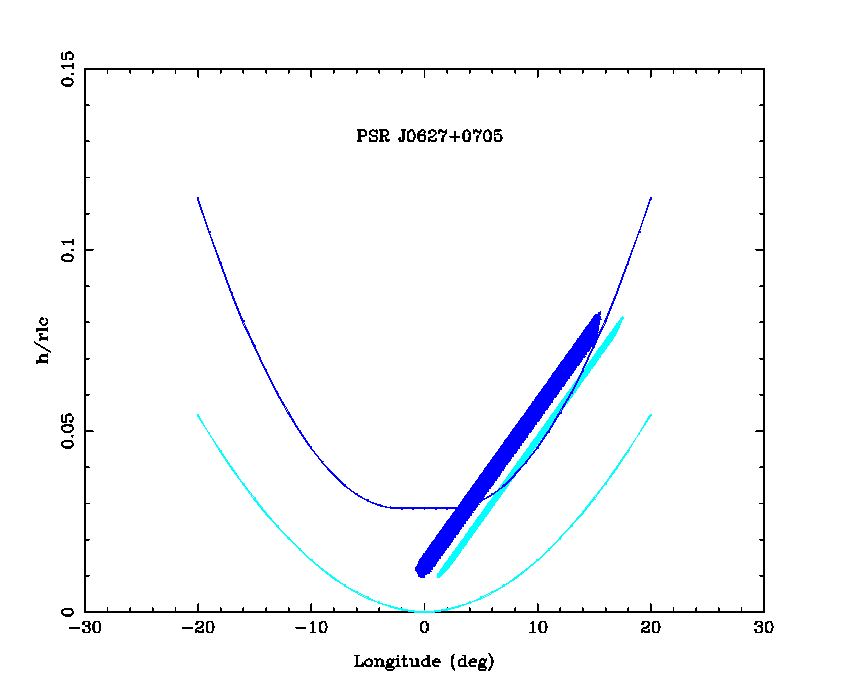} & \includegraphics[width=0.40\textwidth]{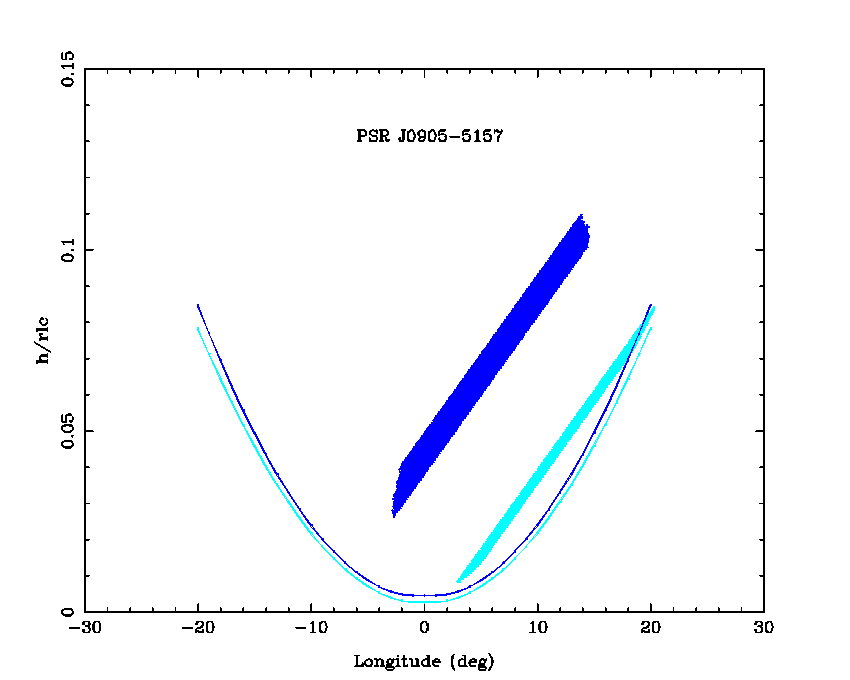} \\
\includegraphics[width=0.40\textwidth]{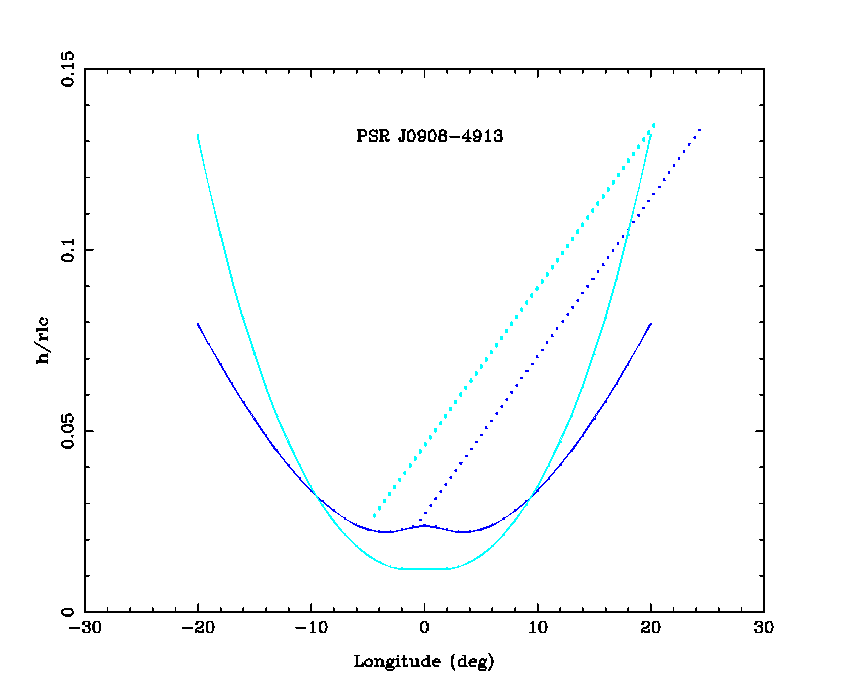} & \includegraphics[width=0.40\textwidth]{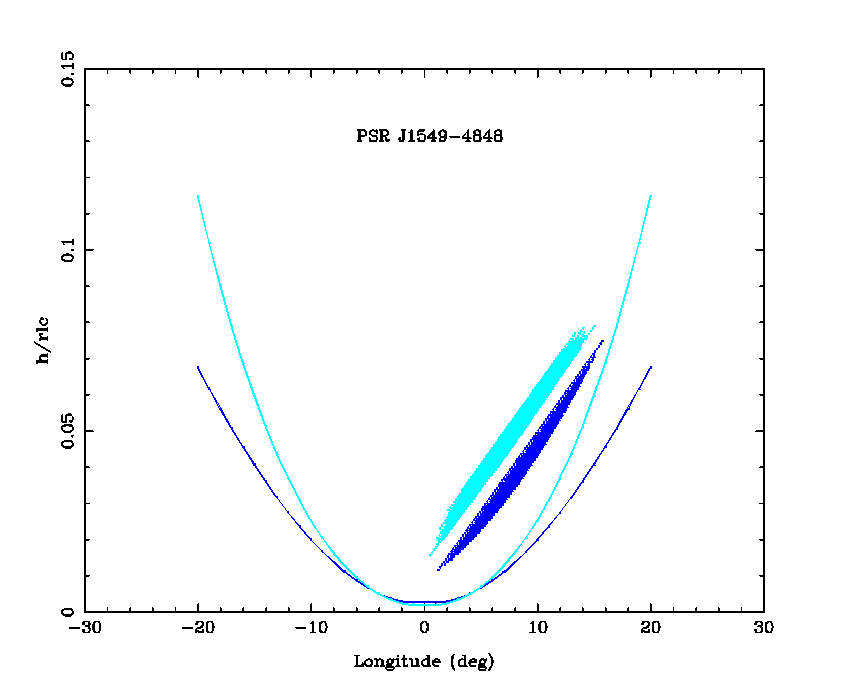} \\
\includegraphics[width=0.40\textwidth]{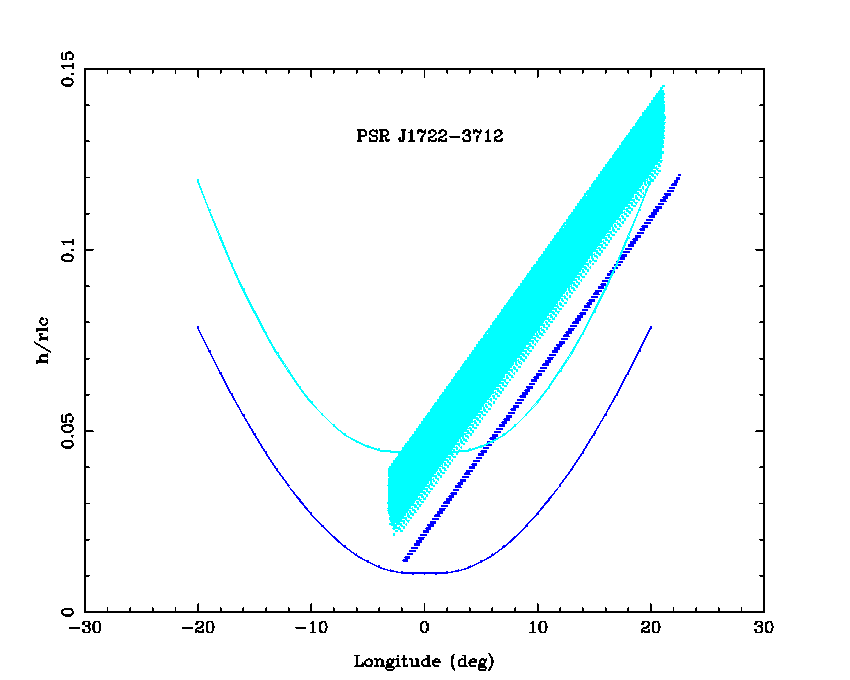} & \includegraphics[width=0.40\textwidth]{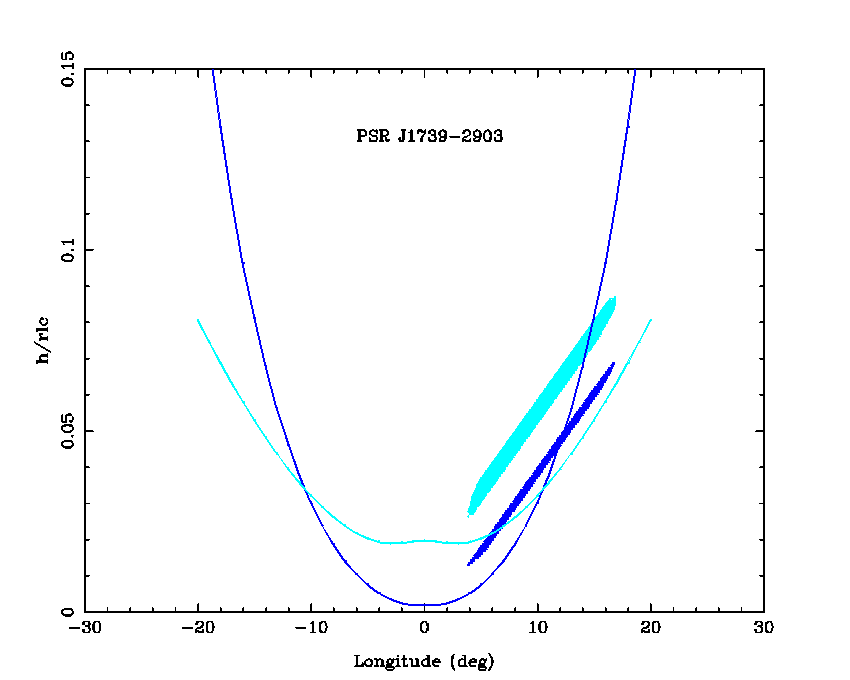} \\
\includegraphics[width=0.40\textwidth]{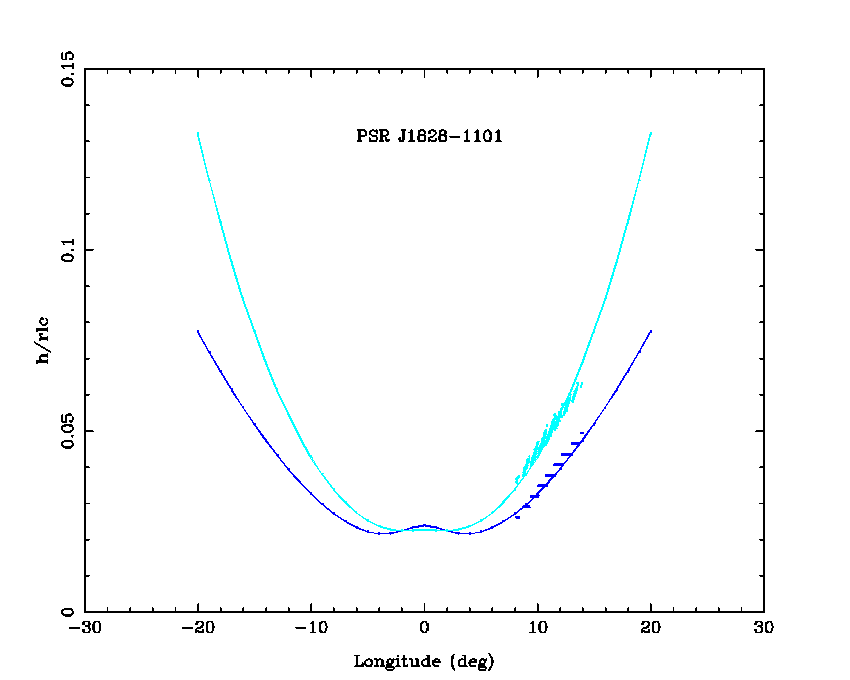} & \includegraphics[width=0.40\textwidth]{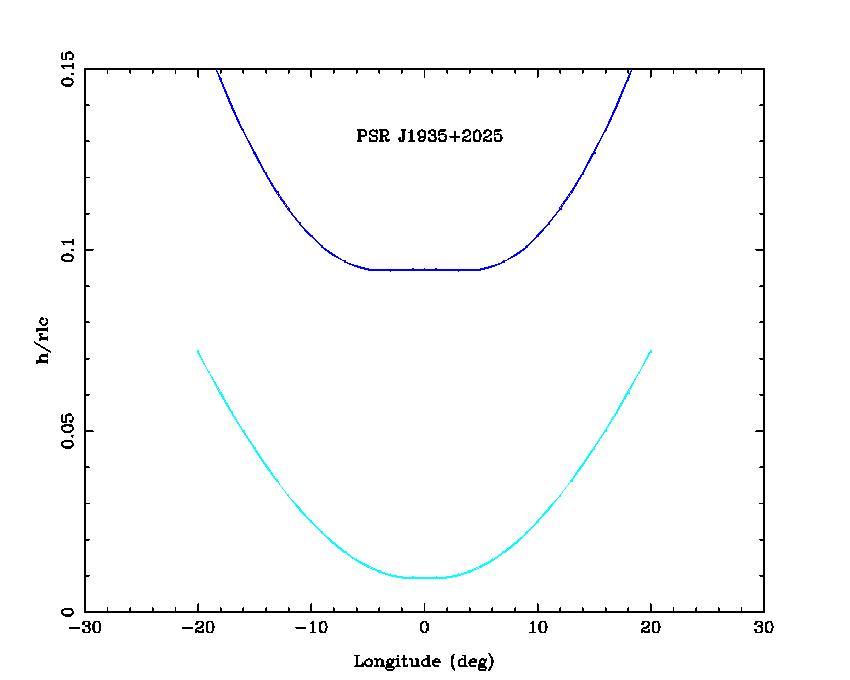} \\
\end{tabular}
\caption{Longitude versus emission height for 8 pulsars. Dark blue denotes main pulse, light blue interpulse. Curves represent the minimum height for the emission to be visible to an Earth-based observer according to \citet{ym14}. Shaded areas are values permitted from the RVM fit and associated errors under the assumption that the emission must occur within the open field line region.}
\label{ymfig}
\end{figure*}

Figure~\ref{ymfig} is a pictorial representation of the values from Table~\ref{beamtab}. For each pulsar, the dark blue area represents possible location of the emission zone of the main pulse in longitude and height. Light blue denotes the interpulse. In addition the minimum height conditions of \citet{ym14} are shown; emission {\bf below} this line is therefore not visible. For two of the pulsars, this minimum height plays a role. For PSR~J1722--3712, the interpulse cannot be centered on the leading edge but must be (at least) towards the beam centre. For PSR~J0607+0705 the main pulse must be substantially shifted towards the trailing edge. In the other five cases the minimum height solution determined by the RVM fitting is consistent with the minimum height permissible by \citet{ym14}.

\section{Discussion}
Table~\ref{bigtable} summarizes the geometry determined for each pulsar. The phase offset, $\Delta$, expressed in degrees of longitude, can be translated into a difference in emission height for main pulse, $h_M$, and interpulse, $h_I$. We find only one case where $\Delta$ is negative, i.e. where the emission height of the main pulse is larger. Three cases are consistent with equal emission height, while for four pulsars, the emission height of the interpulse appears larger. In contrast, we find an equal number of pulsars with an `outer' line-of-sight (los) cut compared to an `inner' los, where an outer los cuts the beam on the far side to the rotation axis with respect to the beam centre given by the magnetic axis (see Table~\ref{circ}).

The filling fraction is given in the last column of Table~\ref{beamtab}. The average filling fraction, given the minimum emission height solutions is 0.57 (excluding PSR~J1935+2025). This is somewhat smaller to the average value of 0.7 found by \citet{jk19} in  their study of the pulse widths of the population as a whole. There appears to be some preference for the emission to be located on the trailing part of the beam. For the minimum height solutions, three are symmetric about the magnetic axis, 10 are mainly on the trailing half and three on the leading half of the beam. In theoretical models, this may be consistent with cyclotron absorption varying across the beam \citep{lm01}. For the maximum height solutions, all the emission is from the trailing half of the beam and the overall filling factor is very low.

We have a large range of possible height solutions for each pulsar from Table~\ref{beamtab} and Figure~\ref{ymfig}. To pick a solution we choose (a) minimum height allowable after taking \citet{ym14} into account (b) symmetry about the magnetic axis and (c) a location on the pulse profile at which the location of the pole is favoured. There are no a-priori reasons for these choices other than (a) trying to maximise the beam filling factor (see \citealt{jk19}) and (b) the non-physical looking nature of the maximum height beam pattern especially given our knowledge of pulsars which show clear symmetrical structure. We discuss the choices for each pulsar in the subsection below.

\begin{figure*}
\begin{tabular}{cc}
\includegraphics[width=7cm]{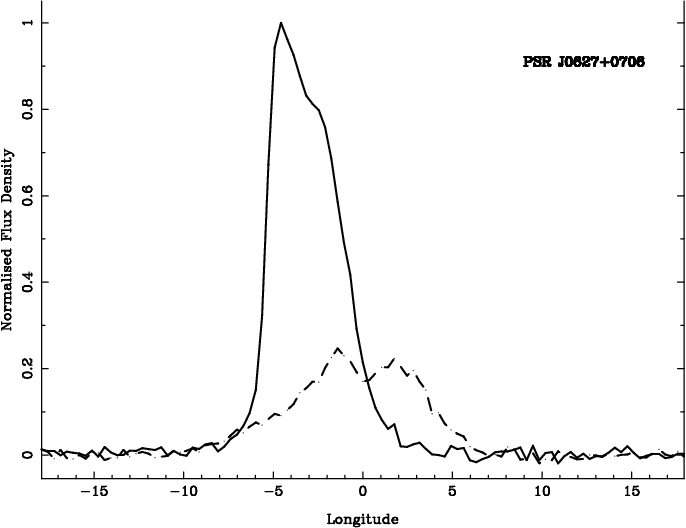} & \includegraphics[width=7cm]{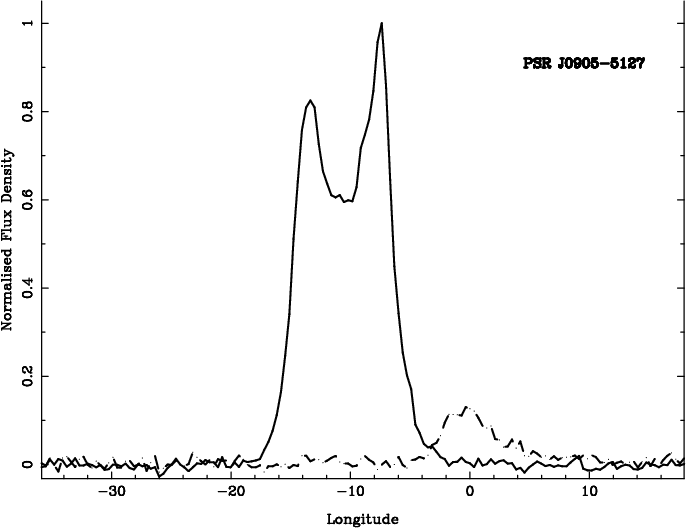} \\
\includegraphics[width=7cm]{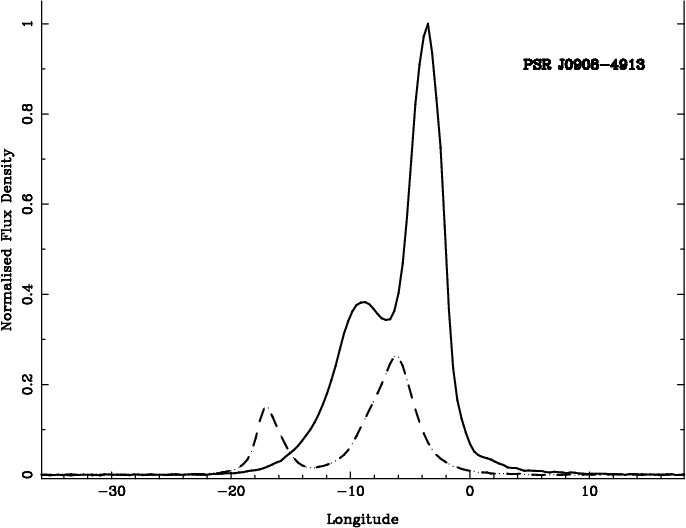} & \includegraphics[width=7cm]{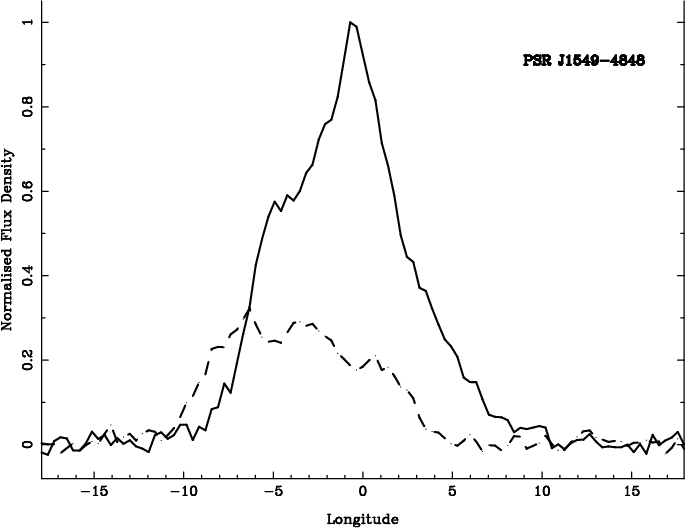} \\
\includegraphics[width=7cm]{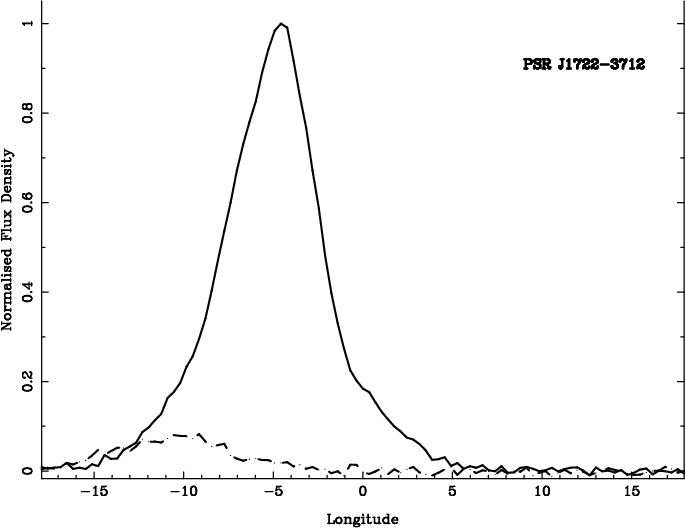} & \includegraphics[width=7cm]{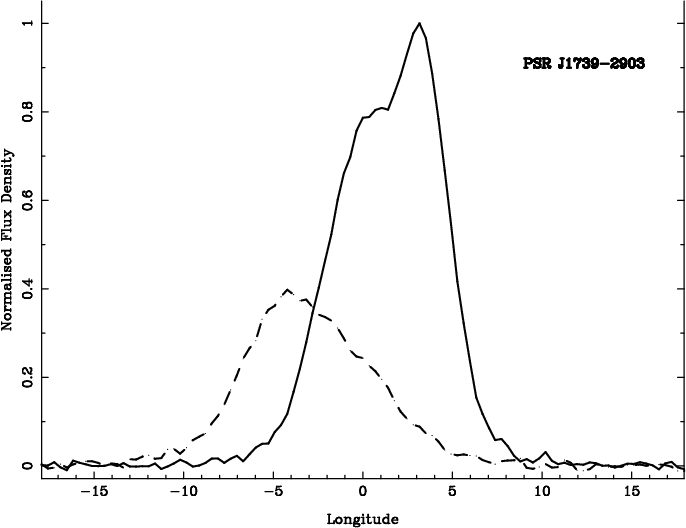} \\
\includegraphics[width=7cm]{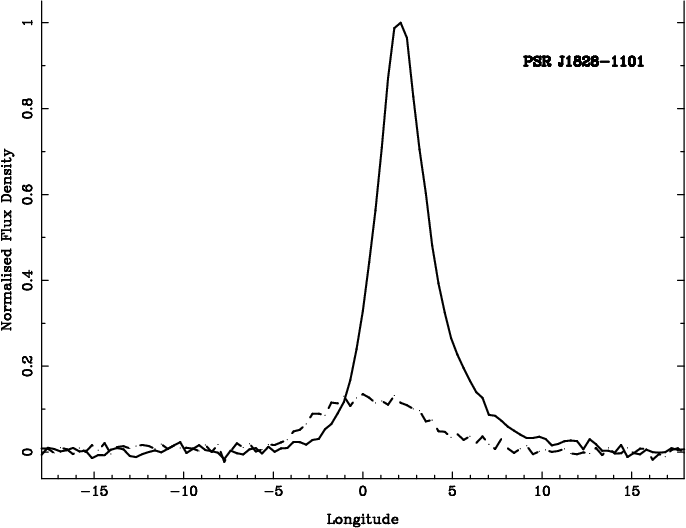} & \includegraphics[width=7cm]{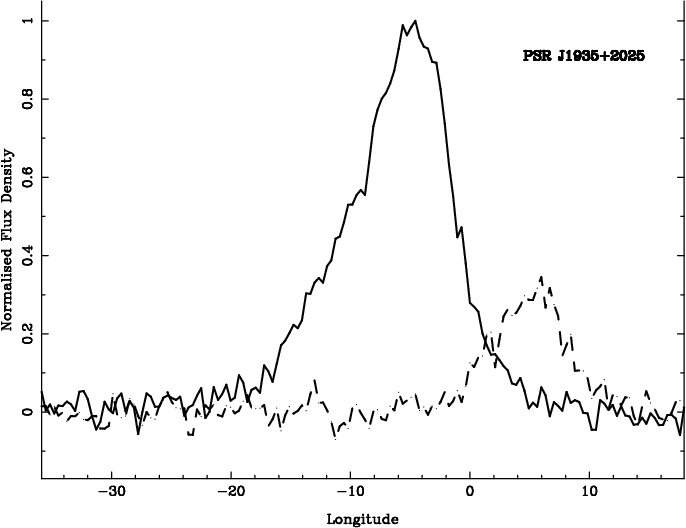} \\
\end{tabular}
\caption{Main (solid line) and interpulse (dashed line) profiles for 8 pulsars, normalised to the peak of the main pulse. Longitude 0\degr\ marks the location of $\phi_0$ for both main and interpulse after taking $\Delta$ into account.}
\label{fig:profiles}
\end{figure*}
\subsection{Individual pulsars}
The total intensity profiles for each pulsar are given in Figure~\ref{fig:profiles} with the main pulse is shown as a solid line and the interpulse as a dashed line. In the figure, the zero longitude marks $\phi_0$ as given by the RVM fits for both main and interpulse (i.e. we have corrected for $\Delta$ and also rotated the interpulse by 180\degr). However, this does not represent the location of the magnetic pole with respect to the profile; the effects of aberration and retardation are not taken into account. Polarization profiles for these pulsars can be found in \citet{jk18} and other references given in the individual pulsar sections below.

\noindent
{\bf PSR~J0627+0705:}
In this pulsar, the main pulse consists of a blended double profile with a moderate degree of linear polarization. The circular polarization is negative in the centre of the profile but positive at the edges. The sign change at the trailing edge is coincident with the magnetic pole location identified in the beam plots. The interpulse also has a double profile, there is a small amount of linear and negative circular polarization, and a clear indication of an orthogonal mode jump near the profile centre. The jump is also required by the RVM fit.

The main pulse has a large value of $\beta$, and this coupled with the pulse width means that $\rho$ and hence $h_M$ must be relatively large. Fixing $h_M$ at 270~km to ensure that the profile fits the beam then in turn means that $h_I$ must be a similar height given the value of $\Delta$. The interpulse emission then also largely fills the beam. The maximum heights are set when the interpulse profile intersects with the trailing edge of the beam at $h_I = 1800$~km. The results are consistent with those shown in Figure~1e of \citet{kjwk10}.

Our preferred solution for this pulsar runs from the minimum height of 270~km for the MP through to 500~km, which places the pole location at the peak of the profile in the MP.

\noindent
{\bf PSR~J0905--5157:}
The main pulse for this pulsar shows a classic double horn profile \citep{jw06} and has a high degree of linear polarization. The circular polarization is positive throughout and shows no evidence for a sign change at the crossing of the magnetic pole. The interpulse is weak with an amorphous pulse profile and virtually no circular polarization. An orthogonal mode jump occurs in the centre of the profile, also required in the RVM fitting.

There is a large difference in the emission heights from the two poles, with the main pulse having a higher $\beta$ and higher $h_{em}$. At the minimum of $h_M$, the interpulse grazes the edge of the beam, whereas the main pulse is more symmetric about the pole.  For the maximum $h_I$ the situation is more extreme with a smaller filling fraction and both profiles emanating from the far trailing edge.

Our preferred solution for this pulsar starts at the minimum height of 580~km for the MP to 720~km which places the location of the pole at the symmetry point of the MP profile.

\noindent
{\bf PSR~J0908--4913:} The pulse profile consists of two components and the linear polarization is high for both main and interpulse. The circular polarization is predominantly positive for the main pulse and negative for the interpulse.

In this pulsar, $\beta$ is similar for the main and interpulse and $\Delta$ is small, implying the emission arises from similar heights from both poles.  The minimum $h_M$ is 130~km, the main pulse is symmetric about the pole and only leading edge emission in the interpulse is seen. For the maximum solution the filling fractions are small and both profiles come from the extreme trailing edge of the beam. Note that the solution given in \citet{kj08} where $h_{em}$ is 230~km lies in between the two cases presented here. The \citet{kj08} solution had the interpulse symmetric about the magnetic pole with the main pulse showing trailing emission only. We take the two symmetry cases to cover our preferred height ranges.

\noindent
{\bf PSR~J1549--4848:} The main pulse shows complex structure in total intensity and linear polarization. The circular polarization is mostly negative throughout with the hint of a sign change at the location of the magnetic pole. The interpulse shows signs of a triple structure with the linear polarization highest in the central component. There is little circular polarization.

The difference in $h_{em}$ between the main and interpulse is large, the main pulse has a smaller $\beta$ and a smaller $h_{em}$ than the interpulse. The minimum height solution implies a symmetrical distribution of emission around the pole for both the MP and the IP. The maximum allowable heights have both profiles on the far trailing edge of the beam. This is consistent with Figure~2d of \citet{kjwk10} who also have symmetrically placed beams.

Our preferred solution therefore takes in the minimum height which results in symmetrical beams and a maximum height of 400~km for the MP which locates the pole at $-6$\degr\ in Figure~\ref{fig:profiles}.

\noindent
{\bf PSR~J1722--3712:} The main pulse is triangular in shape with moderate linear polarization and strong positive circular polarization. The interpulse is only some 10\% of the intensity of the main pulse, is highly linearly polarized but has no circular polarization. There are no signs of orthogonal jumps in either the main or interpulse PA swings.

The interpulse, with a higher $\beta$ also has a higher $h_{em}$. The emission is almost symmetrical about the magnetic pole and has a large filling factor.  This result is similar to that given in \citet{kjwk10}. The maximum height allowed is above 10\% of $r_{lc}$ and shows only trailing edge emission. Our preferred height range places the location of the pole at the peak of the MP and the IP respectively.

\noindent
{\bf PSR~J1739--2903:} The main pulse consists of at least two blended components. There is significant structure in the linear polarization with two orthogonal mode jumps present. Indeed, the RVM fitting requires that two jumps are inserted. The circular polarization is positive. The interpulse shows an amorphous structure with moderate linear and negative circular polarization and a smooth swing of PA.

The interpulse, with its higher $\beta$ also has a somewhat higher emission height than the main pulse. Both the main and interpulse emission are located on the trailing half of the beam even at the minimum height; this becomes more extreme for the maximum height case.  The results in \citet{kjwk10} are somewhat different to ours and they prefer emission which extends beyond the edge of the polar cap.

Although symmetry arguments would place the pole near 0\degr\ for the MP and $-4$\degr\ for the IP, neither of these solutions are permitted without the emission overflowing the beam edges. We are therefore left with choosing the minimum height as our preferred solution which locates the pole at $-6.2$\degr\ on Figure~\ref{fig:profiles}.

\noindent
{\bf PSR~J1828--1101:} The narrow main pulse from this pulsar is $\sim$50\% linearly polarization and no circular polarization. The much weaker interpulse, in contrast has a high degree of linear and negative circular polarization. The PA swing is smooth and unbroken for both main and interpulse.

In this pulsar, $\beta$ is similar for both main and interpulse, and the value of $\Delta$ is small. The observed width of the interpulse is considerably larger than that of the main pulse. $\phi_0$ occurs prior to the main pulse peak again implying that mainly trailing edge emission is seen. \citet{kjwk10} show a symmetric solution at a low $h_{em}$, but in this case they are forced to have emission from considerably outside the polar cap region.

Again therefore, even though symmetry arguments would place the pole near $+2$\degr\ for the MP and 0\degr\ for the IP these solutions are not permitted. We can therefore only choose the minimum height solution which has the pole located at $-7$\degr.

\noindent
{\bf PSR~J1935+2025:} This pulsar has a high degree of linear polarization and strong negative circular polarization in both main and interpulse. The PA swings show no signs of orthogonal mode jumps. There is 190\degr\ separation between the peaks of the main and interpulse emission.

We are unable to obtain a solution for this pulsar which does not involve emission from outside the nominal open field lines. The main pulse has a large $\beta$ {\bf and} a large width, but at the same time has $\phi_0$ relatively close to the peak of the profile. This is paradoxical, as the former implies a large emission height is needed to make $\rho$ large enough to match the large $W_{10}$ yet this also forces the profile beyond the trailing edge of the beam. The interpulse has similar problems as $\phi_0$ occurs before the start of the profile again pushing the emission far onto the trailing edge of the beam.

\subsection{Emission heights}
\begin{figure}
\includegraphics[width=9cm]{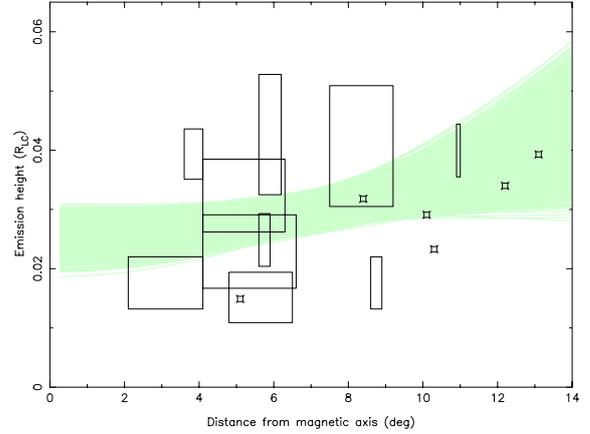}
\caption{Emission height versus distance from the magnetic axis for the pulsars in our sample. The green shaded area indicate parabolic fits to the data using a Monte Carlo method. See text for details.}
\label{parabola}
\end{figure}

In the section above, we identified a range of preferred solutions for the location of the pulse phase of the magnetic axis with respect to both the main pulse and interpulse for five of the pulsars. For the remaining three pulsars only a single solution (minimum height) was found rather than a range. In turn then, these solutions imply a height range and, in conjunction with the geometry determined from the RVM fits, we can compute the distance from the magnetic pole, $d$, to the location of the profile components.
The resulting ranges for the emission heights, in units of light cylinder radius, and distances from the magnetic axis, $d$, are shown as rectangles in Figure~\ref{parabola} for both main and interpulse for five pulsars. For the other three pulsars we indicate the solution with concave squares. Based on the ranges, we see that the emission height tends to be higher at larger distances, $d$. This is consistent with results for individual pulsars as a function of pulse longitude (e.g.~\citealt{gg03}) or magnetic latitude (\citealt{dkl+19}). This can be understood on geometrical grounds for an active emission region bound by dipolar open field lines (e.g.~\citealt{ym14}).

From geometrical considerations, we expect a parabola-like trend for the emission heights as a function of $d$ (c.f.~\citealt{gan04,ym14}). Consequently, we model the data shown in Figure~\ref{parabola} accordingly, using a Monte-Carlo method. From the ten ranges in emission height and distance shown as rectangles, we produce 5000 new data sets, drawing new values from uniform distributions across these ranges.  Each of the new data sets is fitted by a parabola centred on $d=0$. The resulting 5000 fits span the green area indicated in the diagram. 

Being aware that the exact functional dependence of the emission height on $d$ will depend also on the individual geometry of each source, expected to produce a certain scatter, it is interesting to see that a general trend is nevertheless observed.  We point out that the observed trend implies, not unexpectedly, larger beam radii for larger angular distances (ie. $\rho \sim \rm{const.}\times d$). In order to maintain a $\rho\propto 1/\sqrt{P}$ relationship usually observed for non-recycled pulsars (e.g.~\citealt{kwj+94}) this implies that faster spinning pulsars emit further away from the magnetic axis and at higher altitudes \citep{kj07}.

\subsection{Beam structure}
There are a number of striking conclusions which can be drawn from our results. First, there is no evidence for any sort of conal structure in any of these pulsars. Although the profiles appear symmetrical in some cases, it is clear that the symmetry point is not the location of the magnetic pole. This makes conclusions drawn from identification of so-called "core" components particularly problematical (e.g. \citealt{mgr11}). Second, we do not see emission out to the very edge of the beam, either because emission is not produced there or because the emission height is below that necessary for the emission to be visible in the \citet{ym14} formulation. In  either case, evidence from  population studies \citep{kk08} shows that pulsar beams cannot be too under-filled; given our low filling fraction in the longitudinal direction, it therefore seems likely that the beams have a large latitudinal extent such as seen in PSR~J1906+0706 \citep{dkc+13,dkl+19}. Third, there is no obvious relationship between emission height and spin period, because the dominant factor in the height of the emission is its location within the beam. However, faster spinning pulsars may predominantly emit towards the outer edges \citep{jw06,kj07}. Finally, there is a curious dichotomy between the derived emission heights and the (observer based) minimum height of the visible emission derived purely from geometrical considerations \citep{ym14}. Figure~\ref{ymfig} shows this interplay and would seem to imply that pulsars with high $\beta$ are not observable. The implications are beyond the scope of this paper but will be discussed in an upcoming publication.

These points taken together drive us naturally towards an explanation involving an azimuthally limited emission beam which nevertheless produces  radiation over a significant height range along a given field line  (see e.g. Figure~1 in \citealt{dr15}). Furthermore the almost complete absence of frequency evolution in the profile of these pulsars also finds a more natural explanation in the fan-beam model \citep{okj19}.

\begin{table}
\caption{The sign of the circular polarization, $V$, the sign of the
slope of the position angle swing and the sign of $\beta$ for the main
and interpulses for 9 pulsars. The penultimate column denotes whether the
line of sight is an inner (equatorward) or outer (poleward) one. The final
column gives the dominant emission mode according to the model of \citet{bes18}}
\label{circ}
\resizebox{0.50\textwidth}{!}{
\begin{tabular}{lrrrrrrrr}
PSR & V$_M$ & PA$_M$ & $\beta_M$ & V$_I$ & PA$_I$ & $\beta_I$ & los & ox\\
\hline & \vspace{-3mm} \\
0627+0705   & --ve & +ve & --ve &  --ve & +ve & --ve & inner& oo \\
0905--5157  & +ve & +ve & --ve &      & --ve & --ve & inner & x-\\
0908--4913  & +ve & --ve & +ve &  --ve & +ve & --ve & outer & oo \\
1549--4848  & --ve & --ve & +ve &      & +ve & --ve & outer & x- \\
1611--5209  & +ve & +ve & --ve &  --ve & --ve & +ve & outer & xx \\
1722--3712  & +ve & +ve & --ve &      & +ve & --ve & inner  & x- \\
1739--2903  & +ve & +ve & --ve &  --ve & --ve & +ve & outer & xx \\
1828--1101  &     & --ve & +ve &  --ve & +ve & --ve & outer & -o \\
1935+2025   & --ve & +ve & --ve &  --ve & +ve & --ve & inner& oo \\
%1637--4553  & +ve & +ve &  --ve & -ve & outer \\
\end{tabular}}
\end{table}

\subsection{Circular polarisation and geometry}
Table~\ref{circ} shows the comparison between the signs of the circular polarization, $\beta$ and the slope of the position angle swing. A strong correlation is noticeable - if the slope of the PA swing for main and interpulse is the same, then the sign of $V$ is the same. Conversely if the slope of the PA swing changes sign from main to interpulse so does the sign of $V$. This implies that the sign of $V$ depends on the hemisphere of the emission. Indeed such an effect is seen in the precessing pulsar PSR~J1906+0746; when the line of sight crosses the magnetic pole, the sign of $V$ changes \citep{dkl+19}. We recall that in PSR~J0908--4918 the sign of the circular polarization changes in {\em both} the main and interpulse between 1.4 and 3~GHz \citep{kj08}. Indeed there is a strong correlation between the fractional circular polarization in both poles; it is large at low frequencies, both drop to zero at $\sim$2~GHz and then increase again (with opposite sign) at higher frequencies. This maintains our sign of $V$ rule, but as the sign change is not accompanied by an orthogonal mode jump it begs the question as to why the flip of $V$ occurs. We note also the peculiar fact that the change in the sign of $V$ for main pulse and interpulse is observed at exactly the same radio frequency. This does not only strengthen the correlation between main and interpulse, but it could also suggest that a propagation effect depending on the global properties of the magnetosphere above the two poles is responsible.

We note that other observations appear to show in at least some pulsars, a sign change in circular polarization between the leading and trailing parts  of the pulse profile, i.e. at the longitudinal crossing of the magnetic  axis \citep{rr90}. This classic signature is not seen in these interpulse pulsars except for perhaps a hint in PSR~J0627+0706. On the other hand many pulsars have a single sign of circular polarization throughout the pulse profile, and this coupled with the slope of the position angle swing led \citet{bes18} to associate the radio emission with either the X or O modes. If this idea is correct then given the correlation that we see between the  sign of V and that of the PA slope, this implies that the main and the  interpulse emit in the same mode. There is no obvious preference for the X mode over the O mode as also seen in the observations of \citet{jhv+05}. However, it is not clear how the \citet{bes18} model could cope with the sign change in PSR~J0908--4918 described above.

\section{Conclusions}
We have developed a technique for determining the absolute emission heights and the geometry of the main and interpulse emission from pulsars which are close to orthogonal rotators. We apply our technique to a sample of pulsars observed with the Parkes radio telescope. Our conclusions are:
\begin{itemize}
\item[-] In the beam centre, the emission heights are typically less than 400~km in line with the ideas of \citet{mr02} and the simulations of \citet{jk19}.
\item[-] The average filling fraction of the beams in longitude is only 0.57 and there is a strong preference for the visible emission to be located in the trailing part of the beam.
\item[-] There is evidence that the emission heights at the beam edges are significantly higher than those in the beam centre, as expected in the ideas of \citet{gan04} and \citet{ym14} and as also seen observationally by others (e.g. \citealt{gg03,mr02}).
\item[-] The beam patterns that we find point towards a fan-beam like emission structure \citep{wpz+14,dr15,okj19} rather than a `patchy' or `conal' structure.
\item[-] There is a strong correlation between the sign of circular polarization in the main and interpulse profiles; if the slope of the position angle swing is the same in the main and interpulse so is the sign of $V$, if opposite then $V$ also has opposite sign. This implies that the sign of $V$ is related to the hemisphere of the emission.
\end{itemize}

\section*{Acknowledgments}
The Parkes telescope is part of the Australia Telescope National Facility which is funded by the Commonwealth of Australia for operation as a National Facility managed by CSIRO. We thank Axel Jessner for useful comments.

\bibliographystyle{mnras}
\bibliography{ip}
\bsp
\label{lastpage}
\end{document}